\documentclass[apjl]{emulateapj}

\usepackage{graphicx}
\usepackage{amsmath,amssymb}
\usepackage{epstopdf}
\usepackage{color}

\usepackage {pstricks}

\begin{document}

\title{ Synergies between Asteroseismology and Three-dimensional Simulations of Stellar Turbulence}

\author{W. David Arnett\altaffilmark{1} and E. Moravveji\altaffilmark{2} }
\altaffiltext{1}{Steward Observatory, University of Arizona, Tucson, AZ 85721, \email{darnett@as.arizona.edu}
}
\altaffiltext{2}{Institute of Astronomy, KU Leuven, Celestijnenlaan 200D, 3001 Leuven, Belgium}

{\bf DRAFT FROM \today}
 
\begin{abstract}
\noindent Turbulent mixing of chemical elements by convection has fundamental effects on the evolution of stars. 
The standard algorithm at present, mixing-length theory (MLT), is intrinsically local, and must be supplemented 
by extensions with adjustable parameters. 
As a step toward reducing this arbitrariness, we compare asteroseismically inferred internal structures of 
two {\it Kepler} slowly pulsating B stars (SPB's; $M\sim 3.25 M_\odot$) to predictions of 321D turbulence theory, 
based upon well-resolved, truly turbulent three-dimensional simulations \citep{321D,andrea2016} which include 
boundary physics absent from MLT. 
We find promising agreement between the steepness and shapes of the 
theoretically-predicted composition profile outside the convective region in 3D simulations and in asteroseismically 
constrained composition profiles in the best 1D models of the two SPBs.
The structure and motion of the boundary layer,  and the generation of waves, are discussed.
\end{abstract}


\section{Introduction}\label{s-intro}
Observational and computational  advances over the past decade 
have placed us in a 
unique era, in which study of the 
interiors of 
stars can be pursued at much higher precision than before.
The recent observations from space of pulsating stars (through the MOST, CoRoT, {\it Kepler}, K2, and
BRITE-constellation missions), and the planned future missions (like TESS and PLATO) supply precise photometry
of 
stars, with near-continuous time sampling for durations of weeks to years.
Among all targets, those more massive than $\sim1.4\,M_\odot$ have a critical feature in common: 
during their main sequence lives, they harbor a convective core and a radiative envelope.
The gravity (g) modes propagate in the radiative envelope, are reflected from the convective core,
providing valuable information regarding the physical conditions near the boundary.
Thus, g-mode pulsating stars are excellent tests of 
the physics of core convection.


In parallel, two- and three-dimensional (2D/3D) implicit large eddy simulations (ILES) of turbulent convection at 
different evolutionary phases (e.g., \cite{ba94,ns95,fls96,ma06,ma07b,miesch,nsa,viallet2013,321D}, 
and many more) 
have shed light on the 
behavior of stellar convection, and in particular on 
the interface between convective and radiative zones. 
This allows 
the development of non-local time-dependent convective theories 
which are consistent with the 
3D simulations. {\em The numerical data provides closure to the Reynolds-averaged Navier-Stokes 
(RANS) equations \citep{321D}, 
converting numerical experiments into theory. }
This approach is called 321D, and aims to provide alternatives (of increasing sophistication) 
to the classical Mixing Length Theory of convection \citep[MLT,][]{bv58}, 
for use in 
one-dimensional (1D) stellar evolution codes.

Despite this progress in computation, 
simplifications are necessary. Attainable numerical resolution allows numerical Reynolds numbers $Re > 10^4$, which are definitely turbulent, but stars have far higher Reynolds numbers \citep{ropp}. To attain  highly turbulent simulations, only a fraction of the star is computed (a ``box-in-star'' approach) and rotation and magnetic fields are ignored. The simulations extend from the integral scale down into the Richardson-Kolmogorov cascade, and the subgrid dissipation merges with the Kolmogorov ``four-fifths'' law \citep{frisch,amy09vel}.

In addition to the Reynolds number issue, the simulations have negligible radiation diffusion 
(infinite P\'eclet number) because of vigorous neutrino cooling, 
rather than large  but finite P\'eclet  numbers found 
during hydrogen and helium burning \citep{viallet2015}.

These simulations represent turbulent solutions 
to the Navier-Stokes equations, and a step beyond mixing-length theory: 
they can resolve boundaries.
    
It is timely to compare 
the results of observations, asteroseismic modeling 
and 2D/3D simulations
\citep[e.g.][]{conny}. 
Although these are independent approaches, it is possible to 
understand many 
underlying similarities between the state-of-the-art simulations and modeling.
Such synergies shall allow us improve our treatment of turbulent convection in 1D models, and 
account for the convectively-induced mixing in the radiative interior (through overshooting and internal 
gravity waves) in a more consistent way. 

We review  the recent asteroseismic modeling of two {\it Kepler} 
slowly pulsating B stars (SPB's) in \S\ref{s-input}.
In \S\ref{s-compare} we compare the theoretical descriptions in MLT and 321D, region by region.
In \S\ref{s-abund} we compare the shapes of the abundance profiles at the convective boundaries, as inferred from astreroseismology
and from the 3D simulations of O-burning shell in a 23\,$M_\odot$ model \citep{321D} and of C-burning shell in a 15 \ $M_\odot $ model \citep{andrea2015,andrea2016}. Our conclusions are summarized in \S\ref{s-summ}.

\section{Input Models}\label{s-input} 
 \cite{ehsan1, ehsan2} recently did 
 in-depth forward seismic modeling of two SPB stars 
having the richest seismic spectra known so far.
Both \object{KIC\,10526294} \citep{papics1}, and \object{KIC\,7760680} \citep{papics2} are of 
spectral class B8V (3.25\,$M_\odot$), and exhibit long and uninterrupted series of dipole ($\ell=1$) g-modes. 
In both cases, the relative frequency differences between the observations and models is less than 0.5\%.
In addition to placing tight constraints on the extent of overshooting beyond the core, and additional diffusive mixing
in the radiative interior, the authors concluded that the exponentially-decaying diffusive mixing profile for 
core overshoot outperforms the step-function prescription.
Thus, convectively-induced mixing beyond the formal core boundary seems to have a radial dependence, and decays 
outwards.

\begin{figure}[t]
\figurenum{1}
\includegraphics[width=\columnwidth]{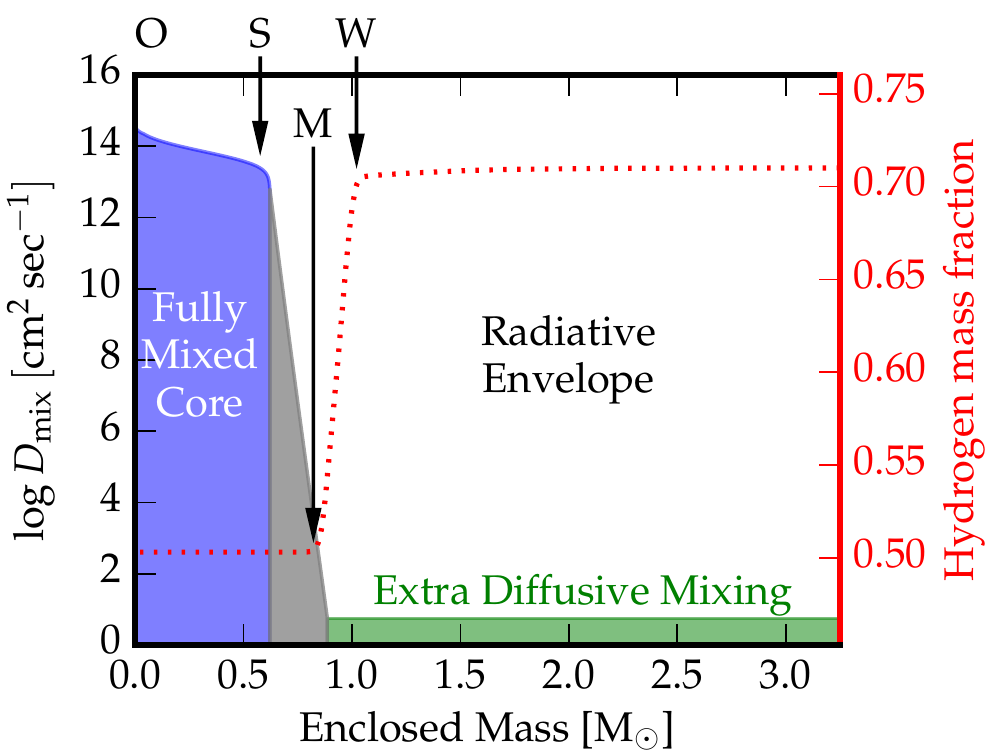}
\caption{Internal mixing in KIC\,7760680. 
The blue region is the formal convective core (MLT), the grey region is the ad-hoc overshoot (braking) layer, and
the green shows the radiative interior which is diffusively mixed at a rate $\log D_{\rm mix}=0.75$ 
(cm$^2$\,sec$^{-1}$).
The origin (O), Schwarzschild boundary (S), bottom of fully mixed region (M) and the top of partially mixed (wave) region (W)
are annotated.
The right ordinate shows the hydrogen abundance profile.
}\label{fig1}
\end{figure}

Figure~\ref{fig1} shows the internal mixing profile (colored regions) of the best 
model of KIC\,7760680, with selected positions in the enclosed mass coordinate labeled O, S, M, and W. 
We will discuss each region, comparing and contrasting the MLT, and a 321D theory based upon 
well-resolved numerical simulations 
(see \cite{321D,andrea2016} and extensive references and discussion therein).

\section{Comparing MLT and 321D}\label{s-compare}

There are four distinct regions in Fig.~\ref{fig1}, which we elaborate below: (1) a Schwarzschild core, (2) a braking (overshoot) region, (3) a composition gradient, and (4) a radiative envelope. 

\subsection{Region OS:  Schwarzschild core}

\subsubsection{MLT}
In Figure~\ref{fig1}, the region extending from the origin at O to S is well mixed; at S the Schwarzschild criterion changes sign, so that within OS buoyancy drives convection. 

For  a composition gradient of zero,  $-\nabla_\mu = \nabla_Y \equiv \Sigma_i X_i/A_i \rightarrow 0$, where $X_i$ and $A_i$ are the mass fraction and mass number of each nucleus $i$.
The Ledoux discriminant,
 $${\cal L} = \Delta \nabla = \nabla - \nabla_{\rm ad} - \nabla_Y, $$ 
reduces to the the Schwarzschild discriminant,
 $${\cal S}  = \nabla - \nabla_{\rm ad},  $$ 
which is positive. 
$\nabla$ and $\nabla_{\rm ad}$ are actual and adiabatic temperature gradients, respectively,  and $\nabla_\mu$ is the corresponding dimensionless gradient in mean molecular weight.
 
 The convective velocity is approximately
\begin{equation}
u^2 \sim g \ell \Delta \nabla >0, \label{eq1}
\end{equation}
where $\ell$ is the free parameter, {\em the mixing length},  
the temperature excess is $\Delta \nabla = \nabla - \nabla_a -\nabla_Y,$ and $g$ is the gravitational acceleration. 
For the turbulent velocity, the MLT as given by Eq.~\ref{eq1} is an adequate 
approximation over region OS, and is the commonly used choice. 
This is a steady-state approximation which may be inferred from the turbulent kinetic energy equation 
\citep{ma07b} or alternatively, from the balance of buoyant acceleration and deceleration by turbulent 
friction ($u|u|/\ell$); see \cite{kolmg41}, \cite{kolmg}, \cite{nathan2014}, and \cite{321D}.

Using the sound speed $s^2 =\gamma PV$, the Mach number ${\cal M} = u/s$, and the pressure scale height $H_P = PV/g$,  we may write Eq.~\ref{eq1} as
$${\cal M}^2 \sim (\ell / H_P) \Delta \nabla .$$ 
For quasi-static stellar evolution, the convective Mach numbers are small (${\cal M} \leq 0.01$). Since $\ell/H_P \sim 1$,  $\Delta \nabla \ll 1$ giving $\nabla \sim \nabla_{\rm ad}$ for a well-mixed convection zone in a stellar interior.

\subsubsection{321D}
Fully turbulent 3D simulations of convection may be represented as solutions to a simple differential equation for either the turbulent kinetic energy, or (as shown here) for  the turbulent velocity $\bf u $ \citep{321D},
\begin{equation}
\partial_t {\bf u} + ({\bf u \cdot \nabla) u} = {\cal B}  - {\cal D} \label{eq2}
\end{equation}
where the buoyancy  term\footnote{For strongly stratified media, there is an additional driving term due to ``pressure dilatation'' \citep{viallet2013}.} is ${\cal B} \approx  g\beta_T \Delta \nabla$  ($\beta_T = (\partial \ln \rho/ \partial \ln  T )_P$ is the thermodynamic factor\footnote{Sometimes denoted $\delta$ or $Q$.} to convert temperature excess to density deficit at constant pressure), and the drag term (chosen to be consistent with the Kolmogorov cascade) is ${\cal D} \approx {\bf u}/\tau$, with the dissipation time $\tau = \ell_d / |u| $. 
Here $\ell_d$ is the dissipation length, a property of the turbulent cascade, and may differ from the mixing length used in MLT.
We average over angles and take the steady state limit, which gives for the radial convective speed $u_r$,
\begin{equation}
 u_r d u_r/dr =  g\beta_T \Delta \nabla  - u_r|u_r|/\ell_d. \label{eq3}
\end{equation}
Away from the convective boundaries the gradient of $u$ is small, and this equation resembles Eq.~\ref{eq1} 
for appropriately chosen $\ell_d$.
Eq.~\ref{eq3} implies a heating rate due to the dissipation of flow at small scales \citep{kolmg41,kolmg}, $\epsilon_K = u^2/\tau = u^3/\ell$, a feature ignored in MLT. {\em This is the frictional cost to  the star for moving enthalpy by turbulence. } In practice this term is small but not negligible.  The 321D algorithm accounts for this additional heating term in the computation of 1D models \citep[see \S4.1.1 in][]{321D}.
 
\subsection{Region SM: Braking (overshoot)} 
\subsubsection{MLT}
The Schwarzschild discriminant, $${\cal S} = \Delta \nabla = \nabla - \nabla_{\rm ad}  $$ changes sign at boundary  S, 
and $\nabla_Y \approx 0$ there, so Eq.~\ref{eq1} implies an imaginary convective speed. To deal with this 
singular 
behavior, different physics is traditionally introduced. A region SM is defined as the ``overshoot'' region, in  which the luminosity is presumably carried entirely by radiative diffusion ($\nabla  = \nabla_{\rm rad}$) and a new algorithm is defined, replacing the  variable $u$ by a new variable, the effective diffusion rate $D_{ov}$; see \cite{ehsan2} for a short and clear discussion. 
Inside SM, ${\nabla_{\rm rad}-\nabla_{\rm ad}\leq\cal S}\leq0$.
The coordinate M designates the layer at which this effective ``convective diffusion'' is no longer able to destroy the composition gradient, so over region SM we still have $\nabla_Y \approx 0$.

\subsubsection{321D}
Because the Ledoux discriminant, $${\cal L} = \Delta \nabla = \nabla - \nabla_{\rm ad} - \nabla_Y $$ changes sign at S, ${\cal B} < 0$, and
the region SM is subjected to buoyant {\em deceleration}. Mixing continues over SM so that all of the region OSM is mixed, even though $\cal L$ is negative over SM. Thus SM is the {\em overshoot region}, in which the flow turns back to complete its overturn. The vector velocity $\bf  u$ has different signs in upflow and downflow; it becomes horizontal at coordinate M, not zero; \citep[see \S3.8 in][]{321D}. The coordinate M is a {\em shear layer.}

Here $\nabla_Y \rightarrow 0$, so we have  $\nabla - \nabla_{\rm ad} < 0$, and $g \ell \Delta \nabla <0$.  
Eq.~\ref{eq1} is not possible.
Near the boundaries the velocity gradient terms dominate over the Kolmogorov term in Eq.~\ref{eq2}, so
\begin{equation}
u_r du_r/dr = d (u_r^2/2)/dr \sim g \beta_T \ell \Delta \nabla. \label{eq4}
\end{equation}
For negative buoyancy ($\Delta \nabla <0$), the buoyant deceleration acts to decrease the radial 
component of the turbulent kinetic energy.
\cite{321D} show that this is essentially the same as defining the boundary by the gradient Richardson 
criterion (${\rm Ri} = N^2/(\partial u /\partial r)^2 > 1 / 4$).

The Schwarzschild criterion is only a linear instability condition, derived by assuming infinitesimal perturbations. The Richardson criterion is a nonlinear condition, which indicates whether the turbulent kinetic energy can overcome the potential energy implied by stable stratification. For a well-mixed  region ${\cal L} \rightarrow {\cal S}$.
Use of $\cal L$ in a stellar code can give 
fictitious boundaries due to small abundance gradients, which are blown away in a 3D fluid dynamic simulation; 
the Richardson criterion insures against these.

The boundary of a convective region has a negligible radial velocity of turbulence; Eq.~\ref{eq4} 
indicates where this occurs, so that by a solution of Eq.~\ref{eq3}, 321D automatically determines where the 
boundary of  convection is.  
This contrasts with MLT for which boundaries are undefined and thus a thorny issue. 


\subsubsection{Radiation diffusion effects}
In order to compare the asteroseismic models from hydrogen burning to simulations from later burning stages, allowance must be made for the difference in strength of radiative diffusion in the two cases. The duration of neutrino-cooled stages is much shorter than the radiative leakage time from the core, while core hydrogen burning takes much longer than its corresponding radiative leakage time. 
In the terminology of fluid mechanics, the P\'eclet number is significantly different in these two cases \citep{viallet2015}, so their flows may be significantly different too. This is especially important for thin layers, for which the radiative diffusion time is shorter.

For oxygen burning and carbon burning, the flow follows a nearly adiabatic trajectory,
having an entropy deficit in the braking region (Fig.~4 in \cite{321D}). If 
radiative diffusion is not negligible, heat will flow into this region of entropy deficit, reducing the strength of the buoyancy braking and thus widening the overshoot layer. This causes the actual gradient to deviate from the adiabatic gradient ($\nabla_{\rm ad}$) and approach the radiative one ($\nabla_{\rm rad}$).
The temperature gradient in the overshooting region is expected to lie between the adiabatic and radiative ones $\nabla_{\rm rad}<\nabla<\nabla_{\rm ad}$ \citep{zahn91,zhang-2016-01}.

Calibration of overshooting algorithms using observations of hydrogen-burning stars may be inadequate for later burning stages because of the differences in P\'eclet number \citep{viallet2015}. This 
could be a problem for helium burning \citep{mosser,jt2015,tom15,ghasemi16}, and will certainly get worse for the neutrino-cooled stages.

\subsection{Region MW: Composition Gradient}\label{MW}
\cite{ehsan1, ehsan2} found it desirable to add an extra diffusive mixing (their $D_{\rm ext}$) beyond the well-mixed region OSM, 
because the agreement between observed and modeled frequencies improve -- in $\chi^2$ sense -- by 
a factor 11. This is an important clue regarding the structure of this region.

The physical basis for extra mixing seems to be at least two-fold. Because coordinate M is a shear layer, 
Kelvin-Helmholtz instabilities  will mix matter above and below  coordinate M. 
In Eq.~\ref{eq3} we saw that the radial velocity could be decelerated at a boundary. As the flow turns, a horizontal velocity $u_h$  must develop \citep{321D}. This variable does not appear in MLT (the ``blob disappears back into the environment"). A finite $u_h$ implies a shear instability may occur.
A necessary condition for instability (due to
Rayleigh and to Fj\o rtoft,  see \S8.2 in \cite{drazin}) is that 
\begin{equation}
d^2 u_h/dr^2 (u_h-u_0) < 0,
\label{fjortoft}
\end{equation} 
somewhere  as we move {\em in radius} through the flow at the boundary.
Here $u_h$ is the horizontal velocity and $u_0$ is that 
velocity at the radius at which $d^2u_h/dr^2 = 0$ (the point of inflection).
While a stably-stratified composition gradient may tend to inhibit the instability, Eq.~\ref{fjortoft} illustrates a basic feature: 
the velocity field drives the instability. 

If this horizontal flow is stable, a composition gradient can be maintained. Turbulent fluctuations may cause the stability criterion to be violated, and thus erode the layer until it is again stable. 
If there is entrainment at the boundary, as the 3D simulations show \citep{ma07b}, then the boundary moves away from the center of  the convection zone.
The result is that the composition gradient will be left near the margin of instability. This gives the {\em sigmoid} shape, found in 3D simulations of oxygen burning \citep{321D} and carbon burning \citep{andrea2016}.

At mass coordinate M the average turbulent velocity in the radial direction is zero. 
Because the flow is turbulent, it is zero only on average, so $\langle u \rangle = 0$, but has a finite rms value, $\langle u^2 \rangle \neq 0$.
Over the regions SM and MW, $\Delta \nabla < 0$, so the Brunt-V\"ais\"al\"a frequency is real. 
Thus, this region can support waves, which the velocity fluctuations necessarily excite. The turbulent fluctuations couple best to g-modes at low Mach numbers \citep{ma07a}.
The mixing induced here is much slower than in the convection region\footnote{Most of the turbulent energy is at large wavelengths (g-modes), so the induced waves are less able to cause mixing than shorter wavelengths might.}, and allows a composition gradient to be sustained over MW. Because the mixing is slow, the composition structure may be 
a combination of previous history and slow modification.

\subsection{Region W: Radiative envelope}

To the extent that nonlinear interactions of waves generate entropy, the circulation theorem implies that slow currents will be induced, leading to more ``extra mixing'' in radiative regions such as those above coordinate W.
In 1D stellar 
evolution, this type of mixing is treated diffusively ($D_{\rm ext}$), and 
asteroseismology can tightly constraint it.

\section{Comparing Abundance Profiles}\label{s-abund}

\begin{figure}[t]
\figurenum{2}
\includegraphics[width=\columnwidth]{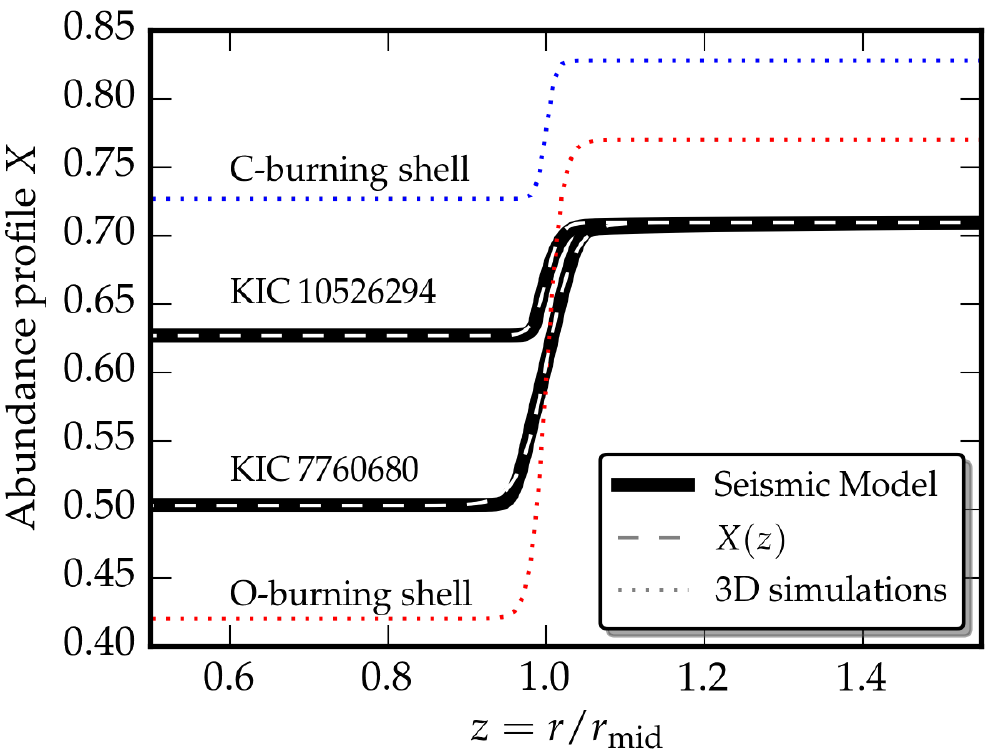}
\caption{Fitting  Eq.\,\ref{e-sigmoid} to the abundance profiles ($X$) of 
KIC\,10526294, and KIC\,7760680. The fit parameters are given in Table\,\ref{t-fit}.
The composition gradients in the 3D simulations \citep{321D,andrea2015} quickly approach a self-similar 
shape of the same form; as shown in \cite{andrea2016}.}
\label{fig2}
\end{figure}

Figure~\ref{fig2} shows the inferred profile of hydrogen in the two {\it Kepler} SPB's. 
This may be well-fitted by a logistic function 
strikingly similar to the composition profile 
shapes resulting from 3D simulations. We choose to fit the abundance profiles $X$ (in mass fraction) to
\begin{equation}
X(z) = \theta + \frac{\phi - \theta}{1 + \exp[-\eta (z-1)]}, \label{e-sigmoid}
\end{equation}
where $z=r/r_{\rm mid}$ is the normalized radius
and $r_{mid}$ is the radius at the mid-point of the abundance gradient.
The steepness of the composition profile is encoded in $\eta$.
At the ```core'', $-\eta(z-1) \gg 1$, and $X(z) \rightarrow \theta$.
At $r=r_{\rm mid} ,$  $X_{\rm mid}= (X_{\rm core} + X_{\rm surf})/2$, 
so $z=1$ and $X_{\rm mid}=0.5(\theta+\phi)$.
At the stellar ``surface'' ($z \gg 1$) and for large $\eta$, $X_{\rm surf}=\phi$.
Convective boundaries in stars are often narrow with respect to radius.
The mathematical form in Eq.~\ref{e-sigmoid} stretches the abundance gradient over the 
whole $z$ axis.
Thus, it is straightforward to fit the sigmoid shape to any abundance profile from stellar models at any
evolutionary phase from 1D or 2D/3D (after temporal and angular averaging).
The normalization used in Eq.~\ref{e-sigmoid} tends to separate shape from absolute scale, so that it may be used for different compositions (burning stages). 

Radiative transfer is negligible for carbon burning and oxygen burning, so the Pecl\'et number is essentially 
infinite \citep{viallet2015}, while radiative transfer does  affect hydrogen burning. 
In the braking region, radial deceleration (turning of the flow) is due to negative buoyancy. At the top of a convection zone, rising matter becomes cooler (has lower entropy) than its surroundings. Radiative transfer tends to heat this cooler
matter, reducing the negative buoyancy (radial braking) so that the matter must continue further before it is turned. This results in a wider braking layer (smaller $\eta$); see discussion in \citep{321D}. 
Boundary layers are narrower for neutrino-cooled stages, so that calibration of boundary widths from 
photon-cooled stages are systematically in error.
Simulations in 3D of H burning are difficult without  artificial scaling of the heating rate \citep{ma07b}. 
Because of negligible  
radiative diffusion in carbon and oxygen burning,  those 
corresponding values for $\eta$ in Table\,\ref{t-fit} should tend to be higher than would be expected in hydrogen burning, as they do. 

Simulations of oxygen burning give behavior similar to carbon burning, but are more complicated to interpret because of a significant initial readjustment of convective shell size, and an episode of ingestion of $^{20}$Ne; see \cite{321D} and references therein to earlier works. Despite this, the  large value of $\eta$  reasonably represents the time- and angle-averaged O-profiles.

\begin{table}[t!]
\caption{Comparison of composition gradients ($\theta,\phi$ and $\eta$ in Eq.\,\ref{e-sigmoid}) 
to H profile of two Kepler SPBs, and to C and O profiles from 3D simulations (resolutions $\sim 1024^3$).}
\begin{center}
\label{t-fit}
\begin{tabular}{lllll}
\hline
Source & Burning & $\theta$ & $\phi$ & $\eta$ \\
\hline \hline 
Asteroseismology: \\
KIC\,10526294 & core H & 0.63 & 0.71 & 112 \\
KIC\,7760680  & core H & 0.50 & 0.71 & 57 \\
\\
3D simulations:\\
upper boundary  & shell C & 0.727 & 0.828 & 184 \\
upper boundary  & shell O  & 0.42 & 0.77 & 103 \\
\hline 
\end{tabular}
\end{center}
\end{table}

In Table~\ref{t-fit} we compare {\em upper} boundaries because
core H burning has only an upper boundary. 
%
The large values of $\eta \gg 1$ imply that the composition boundaries are narrow 
relative to the radius, in all cases. 
The $\eta$ for KIC\,10526294 is roughly twice that of KIC\,7760680.
The former star is very young ($X_{\rm core}=0.63$) with a
steeper composition jump; the latter is more evolved ($X_{\rm core}=0.50$) with broader
$\mu$-gradient zone.
Further investigation of precisely which physical processes determine the parameter $\eta$ is underway (see \S\ref{MW}).

In Fig.~\ref{fig2}, the maximum and minimum mass fractions of fuel are simply a function of the chosen consumption by burning, fixing $\phi$ and $\theta$. The position of the boundary is normalized relative to the center of the gradient $r_{mid}$ (which avoids the important issue of entrainment). What is important here is the narrowness of the composition gradient (the large value of $\eta$) and the shape of the curve joining the high and low fuel abundances, both of which are predicted by the 3D simulations, and independently inferred from asteroseismology, giving an encouraging agreement.

\section{Summary}\label{s-summ}

Asteroseismically-inferred composition gradients are strikingly similar to those found in 
3D simulations; MLT cannot predict boundary structure.
We compare, zone by zone, the predictions of MLT and 321D, from the convective core to the radiative mantle.
Advantages of 321D are its time-dependence, non-locality, incorporation of the 
Kolmogorov cascade and turbulent heating, and resolution of convective boundaries.
The properties of the fully convective regions are similar in 321D and MLT.
Use of the 321D can avoid imaginary convective velocities in the braking regions, which are
related to the development of singularities in boundary layers (\cite{llfm} \S40, \cite{gough77}).
The 321D can provide a continuous description of the convective boundary -- from fully
convective to radiative -- avoiding the awkward patching characteristic of MLT.

The 321D procedure promises a dynamical treatment of the overshooting and wave generation
in stellar models. 
Possible effects  on convective flow from radiative diffusion should be  further explored for regions with moderate P\'eclet numbers \citep{viallet2013}, between existing deep interior and atmospheric simulations.
Better boundaries can provide a possible solution to the behavior of 
convective helium burning cores, which the MLT fails to represent well.

The sigmoid fits to the hydrogen profiles of the two Kepler targets come from 1D (MESA) including
ad-hoc core overshooting and extra diffusive mixing to match the observed g-mode frequencies.
In contrast the simulations of C and O burning shells have no free parameters to adjust. 
Despite dealing with such different burning stages, the similarity shown in Table~\ref{t-fit} 
and Fig.~\ref{fig2} is striking.



\begin{acknowledgements}
Special thanks are due to Simon Campbell and to Andrea Cristini for access to their simulation data prior to publication, and to Raphael Hirschi, Casey Meakin, Cyril Georgy, Maxime Viallet, John Lattanzio and Miro Moc\'ak for helpful discussions.  We thank an anonymous referee who helped to improve the manuscript.
This work was supported in part by the Theoretical Program in Steward Observatory,
and by the People Programme (Marie
Curie Actions) of the European Union's Seventh Framework Programme FP7/2007-2013/ under REA
grant agreement N$^\circ$\,623303 (ASAMBA).
\end{acknowledgements}


\begin{thebibliography}

%
%
%
%
\bibitem[Aerts \& Rogers (2015)]{conny} Aerts, C., \& Rogers, T. M. 2015, \apj, 806, 33

%
%
%
%
%
%
%
%
%
%
\bibitem[Arnett, Meakin \& Young(2009)]{amy09vel} Arnett, W. D., Meakin, C., \& Young, P. A., 2009, \apj, 690, 1715 
\bibitem[Arnett, et al.(2015)]{321D} Arnett, W. D., Meakin, C. A., Viallet, M., Campbell, S. W., Lattanzio, J. C. \& Moc\'ak, M., 2015, \apj, 809, 30 
\bibitem[Arnett \& Meakin(2016)]{ropp} Arnett, W. D. \& Meakin, C. A., 2016, Reports on Progress in Physics, in press

%
%
%
\bibitem[Baz\`{a}n \& Arnett(1994)]{ba94} Baz\`{a}n, G., \& Arnett, D., 
1994, \apj, 433, L41
%
%
%
%
%
%
%
\bibitem[B\"ohm-Vitense(1958)]{bv58} B\"ohm-Vitense, E., 1958, \zap, 46, 108
%
%
%
%
%
%
%
%
%
%
%
%
%
%
%
%
%
%
%
%
%
\bibitem[Constantino, et al.(2015)]{tom15} Constantino, T., Campbell, S., Christensen-Dalsgaard, J., Lattanzio, J., Stello, D., 2015, \mnras, 452, 123
%
%
%
\bibitem[Cristini, et al.(2015)]{andrea2015} Cristini, A., Hirschi, R., Georgy, C., Meakin, C., Arnett, D., \& Viallet, M., 2015, IAUS 307, 459 
%
\bibitem[Cristini, et al.(2016)]{andrea2016} Cristini, A., Hirschi, R., Georgy, C., Meakin, C., Arnett, D., \& Viallet, M., 2016, submitted to \mnras 
%


%
%
%

\bibitem[Drazin(2002)]{drazin}Drazin, P. G., 2002, Introduction to Hydrodynamic Stability, Cambridge University Press, Cambridge, U. K.

%
%
%
%
%
%
\bibitem[Freytag, Ludwig, \& Steffen(1996)]{fls96} Freytag, B., Ludwig, H.-G., \& Steffen, M., 1996, \aap, 313, 497
%
%
\bibitem[Frisch(1995)]{frisch} Frisch, U., 1995, {\it Turbulence}, Cambridge University Press, Cambridge
%
%
\bibitem[Ghasemi, et al.(2016)]{ghasemi16} Ghasemi, H., Moravveji, E., Aerts, C., et al. 2017, \mnras, 465, 1518
%

\bibitem[Gough(1977)]{gough77} Gough, D. O., 1977, 38th Coll., Problems of Stellar Convection (Berlin: Springer), 71, 799
 
%
%
%
%
%
%
%
%
%
%
\bibitem[Kitiashvili, et al.(2016)]{kitiashvili} Kitiashvili, I N., Kosovichev, A. G., Mansour, N. N., \& Wray, A. A. 2016, \apj, 821, 17
\bibitem[Kolmogorov(1941)]{kolmg41} Kolmogorov, A. N., 1941, Dokl.  Akad. Nauk SSSR, 30, 299
\bibitem[Kolmogorov(1962)]{kolmg} Kolmogorov, A. N., 1962, J. Fluid Mech., 13, 82
\bibitem[Landau \& Lifshitz(1959)]{llfm} Landau, L. D. \& Lifshitz, E. M., 1959, Fluid Mechanics,
Pergamon Press, London. 
%
%
%
%
%
%
%
%
%
%
%
%
%
\bibitem[Meakin \& Arnett(2006)]{ma06} Meakin, C. A., \& Arnett, W. D., 2006, \apj, 637, 53
\bibitem[Meakin \& Arnett(2007a)]{ma07a} Meakin, C. A., \& Arnett, W. D., \apj, 665, 690
\bibitem[Meakin \& Arnett(2007b)]{ma07b} Meakin, C. A., \& Arnett, W. D., 2007b, \apj, 667, 448
%
\bibitem[Miesch(2005)]{miesch} Miesch, M. S., 2005, Living Reviews in Solar Physics, 2, 1
%

\bibitem[Moravveji, et al.(2015)]{ehsan1} Moravveji, E.,  Aerts, C., P\'{a}pics, et al. 2015, \aap, 580, 27

\bibitem[Moravveji, et al.(2016)]{ehsan2} Moravveji, E., Townsend, R. H. D., Aerts, C., Mathis, S., 2016, \apj, 823, 130

\bibitem[Mosser et al.(2014)]{mosser} Mosser, B., Benomar, O., Belkacem, K., et al., 2014, \aap, 572, 5

%
%
%
\bibitem[Nordlund \& Stein(1995)]{ns95} Nordlund, A., \& Stein, R., 1995, {\it Stellar Evolution: What Should Be Done},  32nd Li\`ege Int. Astroph. Coll., 32, 75
%
%
\bibitem[Nordlund, Stein, \& Asplund(2009)]{nsa} Nordlund, A., Stein, R., \& Asplund, M., 2009,
  \url{http://www.livingreviews.org/lrsp-2009-2}
%
%
%
\bibitem[Papics et al. (2014)]{papics1} P\'apics, P. I., Moravveji, E., Aerts, C., et al. 2014, \aap, 570, 8

\bibitem[Papics et al. (2015)]{papics2} P\'apics, P. I., Tkachenko, A., Aerts, C., et al. 2015, \apjl, 803, 25
%
%
%
%
%
%
%
\bibitem[Schindler, et al.(2015)]{jt2015} Schindler, J.-T., Green, E. M., \& Arnett, W. D., 2015, \apj, 806, 178
%
%
%
\bibitem[Smith \& Arnett(2014)]{nathan2014} Smith, Nathan, \& Arnett, W. D., 2014, \apj, 785, 82
%
%
%
%
%
%
%
%
%
\bibitem[Viallet, et al.(2013)]{viallet2013} Viallet, M., Meakin, C., Arnett, D., Moc\'ak, M., 2013, \apj, 769, 1

\bibitem[Viallet, et al.(2015)]{viallet2015} Viallet, Maxime, Meakin, C., Prat, V., \& Arnett, D., 2015, \aap, 580, 61 
%
%
%
%
%
%
%
%
%
%
%
\bibitem[Zahn(1991)]{zahn91} Zahn, J.-P. 1991, \aap, 252, 179
\bibitem[{{Zhang}(2016)}]{zhang-2016-01} Zhang, Q.~S. 2016, \apj, 818, 146
\end{thebibliography}
\end{document}